# Weighted least squares estimation by multivariate-dependent weights for linear regression models

Lei Huang, Chengyue Liu*, Li Wang

*Department of Statistics, School of Mathematics, Southwest Jiaotong University, Sichuan, China*

**Abstract**

Multivariate linear regression models often face the problem of heteroscedasticity caused by multiple explanatory variables. The weighted least squares estimation with univariate-dependent weights has limitations in constructing weight functions. Therefore, this paper proposes a multivariate dependent weighted least squares estimation method. By constructing a linear combination of explanatory variables and maximizing their Spearman rank correlation coefficient with the absolute residual value, combined with maximum likelihood method to depict heteroscedasticity, it can comprehensively reflect the trend of variance changes in the random error and improve the accuracy of the model. This paper demonstrates that the optimal linear combination exponent estimator for heteroscedastic volatility obtained by our algorithm possesses consistency and asymptotic normality. In the simulation experiment, three scenarios of heteroscedasticity were designed, and the comparison showed that the proposed method was superior to the univariate-dependent weighting method in parameter estimation and model prediction. In the real data applications, the proposed method was applied to two real-world datasets about consumer spending in China and housing prices in Boston. From the perspectives of MAE, RSE, cross-validation, and fitting performance, its accuracy and stability were verified in terms of model prediction, interval estimation, and generalization ability. Additionally, the proposed method demonstrated relative advantages in fitting data with large fluctuations. This study provides an effective new approach for dealing with heteroscedasticity in multivariate linear regression.

## 1 Introduction

Multivariate linear regression, as a classical statistical analysis method, has been widely applied in fields such as economics [1-4], medicine [5-7], sociology [8-10], and psychology [11-13]. Multivariate linear regression typically requires the random error term to satisfy the Gauss-Markov conditions: mean zero, homoscedasticity, and independence. Homoscedasticity refers to the variance of the random error term remaining constant across all values of the explanatory variables, unaffected by changes in these variables. However, in practical applications, it is common for the variance of the error term to vary with different observed values of the explanatory variables, a phenomenon known as heteroscedasticity. When heteroscedasticity exists, the ordinary least squares (OLS) estimates of regression coefficients remain unbiased but are not the least-variance linear unbiased estimates. This can compromise the significance tests for parameters and regression equations to some extent, reducing the model's predictive accuracy and reliability. Therefore, when conducting multivariate linear regression analysis, it is essential to test for heteroscedasticity and implement appropriate measures to mitigate its adverse effects.

Weighted least squares (WLS) is one of the commonly used methods to overcome the adverse effects of heteroscedasticity. Its fundamental principle is that observations with smaller variance in the error term—i.e., those deviating less from the mean—should be given greater weight. Thus, weights are assigned to observations based on the magnitude of their error variance: smaller variance corresponds to larger weights. This ensures that the informational contribution of each observation is roughly balanced. The weights of observed values are typically set as the inverse of the error variance. However, in practical applications, the specific value of the error variance is unknown and must be estimated. Most textbooks [14-16] still have limitations in weight function selection. For instance, in He Xiaoqun et al.'s Applied Regression Analysis [14], In the chapter discussing how to eliminate the adverse effects of heteroscedasticity using weighted least squares, when estimating the variance function of the error term using the maximum likelihood method, only the explanatory variable most strongly correlated with the variation in the error term variance is selected. This variable is then used to construct a power function as the weighting function for the observation points. Theoretically, this method is only applicable when heteroskedasticity is caused by a single variable. However, in practical scenarios, heteroskedasticity is often influenced by multiple variables. For instance, in the financial sector, Xu Renjie [17] empirically demonstrated that stock price volatility is affected by various factors including corporate fundamentals, market sentiment, macroeconomic conditions, and policy changes. In the medical field, Huang Chunping [18] similarly demonstrated using the ARCH model that the conditional variance of healthcare effectiveness data is influenced by variables including economic, cultural, administrative factors, and fluctuations in healthcare project investment amounts.

On the other hand, many scholars employ generalized least squares (GLS) to estimate parameters and eliminate heteroskedasticity effects, but in practical applications, the covariance matrix of the error terms is often unknown. White proposed heteroskedasticity-consistent covariance matrix estimators (HCCMEs) [19], which subsequently spawned a series of error term covariance matrix estimation methods [20-25], such as HC0, HC1, HC2, HC3, HC4, and HC5. However, these methods are all subject to varying degrees of leverage effects. The OR method proposed by Zhang et al. [26], which combines generalized least squares with orthogonal array techniques, exhibits relatively favorable properties but suffers from certain shortcomings in dependent variable selection and tolerance setting. The local polynomial estimation method proposed by Su et al. [27] offers high precision but faces challenges in bandwidth matrix selection under high-dimensional models. Zhang Heguang's two-stage grouping estimation [28] adapts to high-dimensional models but relies solely on the primary factor causing heteroscedasticity during data grouping, neglecting other contributing factors and resulting in partial sample information loss. Zhang Xiaoqin's new two-stage estimation method [29] can fully utilize all sample information, but its simulation experiments still require the assumption that heteroscedasticity is caused solely by a specific explanatory variable—a condition that is difficult to satisfy in practice.

Overall, heteroskedasticity is often associated with multiple explanatory variables. However, existing methods [14,26,28,29] typically employ only the primary explanatory variable responsible for heteroskedasticity to construct functional relationships when estimating heteroskedasticity. This approach overlooks the connections between heteroskedasticity and other variables, resulting in insufficient exploration of heteroskedasticity information. To address this issue, this paper proposes a multivariate-dependent weighted least squares (MVD-WLS) method: it utilizes Spearman's rank correlation coefficient to assess the degree of correlation between variables and changes in residual absolute values, constructing an optimal linear combination of explanatory variables that maximizes the Spearman rank correlation coefficient with residual absolute values. This linear combination's power function serves as an efficient variance estimator for the error term. The optimal exponent is determined via maximum likelihood estimation, and the reciprocal of this power function is selected as the weighting function for weighted least squares estimation, thereby eliminating heteroscedasticity effects.

Compared to existing methods, the innovations and contributions of this paper lie in the following four aspects. First, in terms of methodological innovation, the proposed method comprehensively considers the relationship between the variance of the error term and each explanatory variable, reasonably reflecting the form of heteroscedasticity and demonstrating strong explanatory power for heteroscedastic variations. Second, in terms of algorithm, this paper integrates modern optimization theory and employs heuristic algorithms to compute model parameters. This approach demonstrates strong adaptability to complex scenarios, coupled with robust global search capabilities and flexibility.Third, regarding simulation experiments, this paper designed three distinct forms of heteroskedasticity variation for multi-angle comparison. When the variance structure of the error term is complex, the correlation between individual explanatory variables and residual magnitudes becomes weak. However, the proposed heteroskedasticity estimation method based on linear combinations of variables can still capture the primary information of error variance variation, making the regression model more closely align with reality. It demonstrates broader adaptability across different sample sizes. Fourth, in empirical analysis, this paper employs two real-world datasets from distinct domains to compare the estimation performance of univariate-dependent weights with multivariate-dependent weights from multiple perspectives. Through model cross-validation, it confirms the limitations of univariate-dependent weights while validating the accuracy and effectiveness of the proposed method.

The main content of the subsequent sections is as follows: Section 1 introduces the multivariate linear regression model with heteroscedasticity and its weighted least squares estimation; Section 2 presents the principles of the proposed MVD-WLS algorithm and provides theoretical proofs for the desirable properties of its estimators; Section 3 and Section 4 conduct simulation experiments and empirical analyses, respectively, comparing the performance of the new and existing methods, demonstrating that the proposed method achieves higher accuracy and stronger adaptability; Finally, Section 5 provides a summary and conclusion for the entire paper.

## 2 Heteroscedastic Model and Weighted Least Squares Estimators

This section primarily reviews multivariate linear regression models with heteroscedasticity and the multivariate weighted least squares estimation of their regression equations.

### 2.1 Heteroscedastic Model

Let the general multivariate linear regression model be:

$$\begin{cases} y_i = \beta_0 + \beta_1 x_{i1} + \beta_2 x_{i2} + \cdots + \beta_p x_{ip} + \varepsilon_i, & i = 1, 2, \cdots, n; \\ E(\varepsilon_i) = 0, \ Var(\varepsilon_i) = \sigma_i^2, \ Cov(\varepsilon_i, \varepsilon_j) = 0, & \forall i \neq j. \end{cases} \quad (1)$$

expressed in matrix form as follows:

$$\begin{cases} \mathbf{Y} = \mathbf{X}\boldsymbol{\beta} + \boldsymbol{\varepsilon}, \\ E(\boldsymbol{\varepsilon}) = \mathbf{O}_n, \quad Cov(\boldsymbol{\varepsilon}) = diag\left(\sigma_1^2, \sigma_2^2, \cdots, \sigma_n^2\right), \end{cases} \quad (2)$$

where $\sigma_1^2, \sigma_2^2, \cdots, \sigma_n^2$ is the variance of the random error, and

$$\mathbf{Y} = \begin{pmatrix} y_1 \\ y_2 \\ \vdots \\ y_n \end{pmatrix}, \quad \mathbf{X} = \begin{pmatrix} 1 & x_{11} & \cdots & x_{1p} \\ 1 & x_{21} & \cdots & x_{2p} \\ 1 & \vdots & \ddots & \vdots \\ 1 & x_{n1} & \cdots & x_{np} \end{pmatrix}, \quad \boldsymbol{\beta} = \begin{pmatrix} \beta_1 \\ \beta_2 \\ \vdots \\ \beta_p \end{pmatrix}, \quad \boldsymbol{\varepsilon} = \begin{pmatrix} \varepsilon_1 \\ \varepsilon_2 \\ \vdots \\ \varepsilon_n \end{pmatrix}, \quad \mathbf{O}_n = \begin{pmatrix} 0 \\ 0 \\ \vdots \\ 0 \end{pmatrix}.$$

If the variances of the random error terms are equal, the model is said to have homoscedasticity and satisfy the Gauss Markov condition. Equations (1) and (2) are called classical multivariate linear regression models. If $\sigma_1^2 \neq \sigma_2^2 \neq \cdots \neq \sigma_n^2$, then the model is said to have heteroscedasticity, and correspondingly, equations (1) and (2) are referred to as heteroscedasticity models [29].

## 2.2 Multivariate Weighted Least Squares Estimation

For a general multivariate linear regression model, when the random error exhibits heteroscedasticity, the weighted sum of squares is:

$$Q_w(\boldsymbol{\beta}) = (\mathbf{Y} - \mathbf{X}\boldsymbol{\beta})^T \mathbf{W}(\mathbf{Y} - \mathbf{X}\boldsymbol{\beta}),$$

by minimizing the weighted sum of squared deviations, the matrix expression for the weighted least squares estimate is obtained:

$$\hat{\boldsymbol{\beta}}_w = \left(\mathbf{X}^T \mathbf{W} \mathbf{X}\right)^{-1} \mathbf{X}^T \mathbf{W} \mathbf{Y}, \quad (3)$$

where $\mathbf{W} = diag\{w_1, w_2, \cdots, w_n\}$, $w_i = \dfrac{1}{\sigma_i^2}$, $w_i$ is typically referred to as the weight function, and $\mathbf{W}$ as the weight function matrix.

## 3 Weighted least squares estimation by multivariate-dependent weights

Before performing multivariate weighted least squares estimation using Equation (3), the form of variance change in the error term must first be specified. Since the form of error variance is typically unknown in practical applications, it requires estimation. This section first introduces the method from He Xiaoqun's work [14], which involves selecting the explanatory variable most strongly correlated with the variance of the error term to construct a weighting function. Heteroscedasticity is then estimated using maximum likelihood estimation. Subsequently, the focus shifts to the multivariate dependency-weighted least squares method proposed in this paper.

### 3.1 Univariate-dependent weighted least squares estimation

Construct a weight function using explanatory variable $x_j$, which is most strongly associated

with changes in the variance of the random error, from among all explanatory variables:

$$w_i = x_j^{-m}, \tag{4}$$

for the selection of $x_j$, simply calculate the Spearman rank correlation coefficient between each explanatory variable and the absolute value of the ordinary residual. Select the explanatory variable with the highest correlation coefficient to construct the weight function.

To determine the optimal value of the exponent m, the log-likelihood statistic is used as an indicator of the regression equation's quality. The log-likelihood values corresponding to different m values are calculated, and the m value that maximizes the log-likelihood is identified as the exponent for the weighting function. Substituting this value into equation (4) allows the regression equation to be obtained through weighted least squares estimation.

## 3.2 Multivariate-dependent weighted least squares estimation

In practical applications, heteroskedasticity is often caused by multiple explanatory variables acting together. Estimating heteroskedasticity using only a single explanatory variable ignores the influence of other variables, resulting in the loss of partial sample information and leading to inaccurate estimates of heteroskedasticity. This approach performs particularly poorly when the form of heteroskedasticity variation is complex. Therefore, this paper proposes improving the maximum likelihood estimation method for heteroskedasticity by constructing weighted linear combinations of the explanatory variables.

For the multivariate heteroscedasticity model with p explanatory variables represented by Equation (1), we first apply OLS:

$$\hat{\boldsymbol{\beta}} = (\mathbf{X}^T\mathbf{X})^{-1}\mathbf{X}^T\mathbf{Y},$$

obtained the parameter estimate $\hat{\beta}_0, \hat{\beta}_1, \cdots, \hat{\beta}_p$, then the preliminary regression fit value for the dependent variable $y_i$ $(i=1,2,\cdots,n)$ is:

$$\hat{y}_i = \hat{\beta}_0 + \hat{\beta}_1 x_{i1} + \hat{\beta}_2 x_{i2} + \cdots + \hat{\beta}_p x_{ip},$$

calculate the residuals for $y_i$:

$$e_i = y_i - \hat{y}_i.$$

Constructing a linear combination of explanatory variables:

$$x_i^* = k_1 x_{i1} + k_2 x_{i2} + \cdots + k_p x_{ip}, \tag{5}$$

let $R_i$ denote the rank of $x_i^*$ sorted from smallest to largest, and $E_i$ denote the rank of the absolute value of the ordinary residual $|e_i|$. Then the rank difference between $x_i^*$ and $|e_i|$ is:

$$d_i = |R_i - E_i|,$$

by maximizing the absolute value of the Spearman rank correlation coefficient between $x_i^*$ and $|e_i|$, the optimal weight coefficient $\hat{k}_1, \hat{k}_2, \cdots, \hat{k}_p$ for the linear combination of explanatory variables is obtained:

$$\mathbf{k} = \left(\hat{k}_1, \hat{k}_2, \cdots, \hat{k}_p\right)^T = \underset{k_1, k_2, \cdots, k_p}{\arg\max} \left\{|r_s|\right\} = \underset{k_1, k_2, \cdots, k_p}{\arg\max} \left\{\left|1 - \frac{6}{n(n^2-1)}\sum_{i=1}^{n} d_i^2\right|\right\}. \tag{6}$$

Construct the variance equation for the random error:
$$\sigma_i^2 = \sigma^2 (\mathbf{x}_i^T \mathbf{k})^m \tag{7}$$
where m is the parameter to be estimated, under the assumption of normality, the log-likelihood function of the heteroscedastic multivariate linear regression model is:

$$l(\boldsymbol{\beta}, \sigma^2, m) = -\frac{n}{2}\ln(2\pi) - \frac{n}{2}\ln(\sigma^2) - \frac{m}{2}\sum_{i=1}^{n}\ln(\mathbf{x}_i^T \mathbf{k}) - \frac{1}{2\sigma^2}\sum_{i=1}^{n}\frac{(y_i - \mathbf{x}_i^T \boldsymbol{\beta})}{(\mathbf{x}_i^T \mathbf{k})^m},$$

let $w_i = \mathbf{x}_i^T \mathbf{k}$, from the above, each $w_i$ is known. To estimate the parameters $\boldsymbol{\beta}, \sigma^2, m$, taking the first-order partial derivatives of the log-likelihood function with respect to parameters $\boldsymbol{\beta}, \sigma^2, m$ yields:

$$\frac{\partial l}{\partial \boldsymbol{\beta}} = \frac{1}{\sigma^2}\sum_{i=1}^{n}\frac{(y_i - \mathbf{x}_i^T \boldsymbol{\beta})}{w_i^m}\mathbf{x}_i,$$

$$\frac{\partial l}{\partial \sigma^2} = -\frac{n}{2\sigma^2} + \frac{1}{2(\sigma^2)^2}\sum_{i=1}^{n}\frac{(y_i - \mathbf{x}_i^T \boldsymbol{\beta})^2}{w_i^m},$$

$$\frac{\partial l}{\partial m} = -\frac{1}{2}\sum_{i=1}^{n}\ln(w_i) + \frac{1}{2\sigma^2}\sum_{i=1}^{n}\frac{(y_i - \mathbf{x}_i^T \boldsymbol{\beta})^2 \ln(w_i)}{w_i^m},$$

treating parameter m as a fixed value,, the likelihood equation for parameter $\boldsymbol{\beta}$ *and* $\sigma^2$ yields:

$$\hat{\boldsymbol{\beta}}(m) = (\sum_{i=1}^{n} w_i^{-m}\mathbf{x}_i \mathbf{x}_i^T)(\sum_{i=1}^{n} w_i^{-m}\mathbf{x}_i y_i),$$

$$\hat{\sigma}^2(m) = \frac{1}{n}\sum_{i=1}^{n}\frac{(y_i - \mathbf{x}_i^T \hat{\boldsymbol{\beta}}(m))}{w_i^m},$$

substituting $\hat{\boldsymbol{\beta}}(m)$ and $\hat{\sigma}^2(m)$ into the likelihood equation for m, the maximum likelihood estimate of m is obtained by solving the following equation:

$$\frac{\sum_{i=1}^{n}\frac{\left(y_i - \mathbf{x}_i^T \hat{\boldsymbol{\beta}}(m)\right)^2 \ln(w_i)}{w_i^m}}{\sum_{i=1}^{n}\frac{\left(y_i - \mathbf{x}_i^T \hat{\boldsymbol{\beta}}(m)\right)^2}{w_i^m}} = \frac{1}{n}\sum_{i=1}^{n}\ln(w_i).$$

The above outlines the fundamental principles of multivariate-dependent weighted least squares proposed in this paper. Below is a summary of the algorithm's steps:

**Algorithm: Multivariate-dependent Weighted Least Squares (MVD-WLS)**

**Step 1:** Using the sample data, calculate the preliminary regression equation based on OLS and compute the absolute value of the ordinary residuals $|e_i|$.

**Step 2:** Constructing linear combinations of explanatory variables:
$$x_i^* = k_1 x_{i1} + k_2 x_{i2} + \cdots + k_p x_{ip},$$
$$d_i = |R_i - E_i|,$$
$$\mathbf{k} = \left(\hat{k}_1, \hat{k}_2, \cdots, \hat{k}_p\right)^T = \arg\max_{k_1, k_2, \cdots, k_p}\{|r_s|\} = \arg\max_{k_1, k_2, \cdots, k_p}\left\{\left|1 - \frac{6}{n(n^2-1)}\sum_{i=1}^{n} d_i^2\right|\right\}.$$

**Step 3:** Estimating the optimal exponent m using the maximum likelihood method:

Initialize $m^{(0)} = 0$.

Iteration steps: for $k = 0, 1, 2, \cdots$, do

| Calculate weights: $w_i^{m^{(k)}} = (\mathbf{x}_i^T \mathbf{k})^{m^{(k)}}$;

| Calculate the weighted least squares estimator: $\hat{\boldsymbol{\beta}}^{(k)}(m) = (\sum_{i=1}^{n} w_i^{-m^{(k)}} \mathbf{x}_i \mathbf{x}_i^T)(\sum_{i=1}^{n} w_i^{-m^{(k)}} \mathbf{x}_i y_i)$;

| Calculate residuals: $r_i^{(k)} = y_i - \mathbf{x}_i^T \hat{\boldsymbol{\beta}}^{(k)}(m)$;

| Update m, by solving the equation: $\sum_{i=1}^{n} \frac{r_i^{(k)2} \ln(w_i)}{w_i^m} = \overline{\ln w} \sum_{i=1}^{n} \frac{r_i^{(k)2}}{w_i^m}$.

Convergence condition: When $\left| m^{(k+1)} - m^{(k)} \right| < \varepsilon$, stop iteration.

**Step 4:** Using the the optimal exponent $\hat{m}$ and optimal combination coefficient $\mathbf{k}$ to constructing the weighting function $\omega_i = (\mathbf{x}_i^T \mathbf{k})^{-\hat{m}}$, then using WLS, the final regression equation is obtained.

### 3.3 Theoretical Property

The subsequent sections of this chapter will demonstrate that the optimal linear combination exponent estimator $\hat{m}$ for heteroscedasticity fluctuations, derived from the MVD-WLS algorithm, possesses convergence properties and asymptotic normality.

When parameter $\boldsymbol{\beta}$ and $\sigma^2$ takes its true value, let the probability density function of the population be:

$$p(y; m, \boldsymbol{\beta}_0, \sigma_0^2) = \frac{1}{\sqrt{2\pi\sigma_0^2 w^m}} \exp\left\{ -\frac{(y - \mathbf{x}^T \boldsymbol{\beta}_0)^2}{2\sigma_0^2 w^m} \right\},$$

sample $(x_i, y_i), i = 1, 2, \cdots, n$ is drawn from this population. Let $l(m, \boldsymbol{\beta}_0, \sigma_0^2; y) = \sum_{i=1}^{n} \ln p(y_i; m, \boldsymbol{\beta}_0, \sigma_0^2)$, to ensure the rigor of the verification process, the following assumptions are made:

***Assumption 1.*** At least one $w_i \neq 1$ exists, and $\{w_i\}$ is not entirely equal, $i = 1, \cdots n$.

***Assumption 2.*** There exist constants $0 < c_w < C_w < \infty$ such that $c_w < w_i < C_w$ hold for $i = 1, \cdots n$.

***Assumption 3.*** The parameter space $\Theta$ for m is an open interval.

Assumption 1 indicates that since the research subject is a heteroscedastic model, if all $w_i$ are equal, then according to the variance equation in (7), the heteroscedastic model will degenerate into an isoscedastic model, not all $w_i$ are equal ensures the model's validity, while the existence of at least one $w_i \neq 1$ guarantees the validity of the following Lemma 1.

Assumption 2 indicates that $w_i^m$ does not tend toward zero nor exhibit explosive growth. As shown by Equation (6), this method focuses more on the relative magnitudes of the components of the combination coefficient vector $\mathbf{k}$. By appropriately adjusting the multiplicative factors, Assumption 2 can be readily satisfied, thereby ensuring the validity of the following Lemma 2.

Based on the above assumptions, we first prove several preliminary lemmas:

**Lemma 1:** Suppose Assumptions 1 hold, the density function $p(y;m,\boldsymbol{\beta}_0,\sigma_0^2)$ of the population is identifiable with respect to the parameter m, i.e.:

For $\forall m_1 \neq m_2$, $\{y: p(y;m_1,\boldsymbol{\beta}_0,\sigma_0^2) \neq p(y;m_2,\boldsymbol{\beta}_0,\sigma_0^2)\}$ is not a zero measure set.

**Proof**

For $\forall m_1 \neq m_2$, we assume: $p(y;m_1,\boldsymbol{\beta}_0,\sigma_0^2) = p(y;m_2,\boldsymbol{\beta}_0,\sigma_0^2)$ for all $y \in R$, taking the logarithm of both sides of the this equation yields:

$$\frac{(y-\mathbf{x}^T\boldsymbol{\beta}_0)^2}{2\sigma_0^2}\left(\frac{1}{w^{m_2}} - \frac{1}{w^{m_1}}\right) + \frac{1}{2}(m_2 - m_1)\ln w = 0. \tag{8}$$

Now consider $y = \mathbf{x}^T\boldsymbol{\beta}_0$, equation (8) holds for all $y \in R$ if and only if $\ln w = 0$, this contradicts Assumptions 1, thus the original hypothesis does not hold. Furthermore, since a quadratic equation has at most two roots, there are at most finitely many values of y satisfying equation (8). Since the Lebesgue measure of a finite set is zero, we have:

$$m(\{y: y \in \mathbb{R}\}) = m\left(\{y: p(y;m_1,\boldsymbol{\beta}_0,\sigma_0^2) \neq p(y;m_2,\boldsymbol{\beta}_0,\sigma_0^2)\} \cup \{y: p(y;m_1,\boldsymbol{\beta}_0,\sigma_0^2) = p(y;m_2,\boldsymbol{\beta}_0,\sigma_0^2)\}\right)$$

$$= m\left(\{y: p(y;m_1,\boldsymbol{\beta}_0,\sigma_0^2) \neq p(y;m_2,\boldsymbol{\beta}_0,\sigma_0^2)\}\right) + m\left(\{y: p(y;m_1,\boldsymbol{\beta}_0,\sigma_0^2) = p(y;m_2,\boldsymbol{\beta}_0,\sigma_0^2)\}\right)$$

$$= m\left(\{y: p(y;m_1,\boldsymbol{\beta}_0,\sigma_0^2) \neq p(y;m_2,\boldsymbol{\beta}_0,\sigma_0^2)\}\right) + 0 = +\infty > 0.$$

Therefore, the overall density function $p(y;m,\boldsymbol{\beta}_0,\sigma_0^2)$ is identifiable with respect to the parameter m.

***Remark 1.*** Lemma 1 guarantees that the probability density function of the population is identifiable with respect to parameter m. In proving the consistency of the maximum likelihood estimate, Jensen's inequality is required to demonstrate that the true value of the parameter is the maximum point of the expected log-likelihood function. Identifiability ensures the uniqueness of this maximum point.

**Theorem 1:** The optimal linear combination exponent estimator $\hat{m}$ for heteroscedasticity obtained by the MVD-WLS algorithm is consistent.

**Proof**

By Lemma 1, we know that $p(y;m,\boldsymbol{\beta}_0,\sigma_0^2)$ is recognizable with respect to the parameter m. For $\forall m' \neq m$, by Jensen's inequality:

$$E_m\left(\ln \frac{p(y;m',\boldsymbol{\beta}_0,\sigma_0^2)}{p(y;m,\boldsymbol{\beta}_0,\sigma_0^2)}\right) < \ln\left[E_m\left(\frac{p(y;m',\boldsymbol{\beta}_0,\sigma_0^2)}{p(y;m,\boldsymbol{\beta}_0,\sigma_0^2)}\right)\right] = \ln 1 = 0.$$

Let the actual value of parameter $m$ be $m_0$, for sufficiently small $\delta > 0$, satisfying $(m_0 - \delta, m_0 + \delta) \subset \Theta$ and $E_{m_0}\left(\ln \frac{p(y;m_0-\delta,\boldsymbol{\beta}_0,\sigma_0^2)}{p(y;m_0,\boldsymbol{\beta}_0,\sigma_0^2)}\right) < 0$, $E_{m_0}\left(\ln \frac{p(y;m_0+\delta,\boldsymbol{\beta}_0,\sigma_0^2)}{p(y;m_0,\boldsymbol{\beta}_0,\sigma_0^2)}\right) < 0$.

According to the Law of Large Numbers, when $n \to \infty$:

$$\frac{1}{n}\left[l(m_0-\delta,\boldsymbol{\beta}_0,\sigma_0^2;y) - l(m_0,\boldsymbol{\beta}_0,\sigma_0^2;y)\right]$$

$$= \frac{1}{n}\left[\sum_{i=1}^n \ln p(y_i;m_0-\delta,\boldsymbol{\beta}_0,\sigma_0^2) - \sum_{i=1}^n \ln p(y_i;m_0,\boldsymbol{\beta}_0,\sigma_0^2)\right]$$

$$= \frac{1}{n}\left[\sum_{i=1}^n \ln \frac{p(y_i;m_0-\delta,\boldsymbol{\beta}_0,\sigma_0^2)}{p(y_i;m_0,\boldsymbol{\beta}_0,\sigma_0^2)}\right] \xrightarrow{a.s.} E_{m_0}\left(\ln \frac{p(y;m_0-\delta,\boldsymbol{\beta}_0,\sigma_0^2)}{p(y;m_0,\boldsymbol{\beta}_0,\sigma_0^2)}\right) < 0,$$

similarly, we can obtain:

$$\frac{1}{n}\left[l(m_0+\delta,\boldsymbol{\beta}_0,\sigma_0^2;y)-l(m_0,\boldsymbol{\beta}_0,\sigma_0^2;y)\right] \xrightarrow{a.s.} E_{m_0}\left(\ln\frac{p(y;m_0+\delta,\boldsymbol{\beta}_0,\sigma_0^2)}{p(y;m_0,\boldsymbol{\beta}_0,\sigma_0^2)}\right)<0,$$

since $l(m,\boldsymbol{\beta}_0,\sigma_0^2;y)$ is continuous on $[m_0-\delta,m_0+\delta]$, there must exist a local maximum point, denoted as $\hat{m}$.

Consider the differentiability of $l(m,\boldsymbol{\beta}_0,\sigma_0^2;y)=\sum_{i=1}^n \ln p(y_i;m,\boldsymbol{\beta}_0,\sigma_0^2)$ on $\Theta$:

$$\frac{\partial \ln p(y;m,\boldsymbol{\beta}_0,\sigma_0^2)}{\partial m}=-\frac{1}{2}\ln w+\frac{(y-\mathbf{x}^T\boldsymbol{\beta}_0)^2 \ln w}{2\sigma_0^2 w^m},$$

it is easy to see that $\ln p(y;m,\boldsymbol{\beta}_0,\sigma_0^2)$ is differentiable on $\Theta$, then $l(m,\boldsymbol{\beta}_0,\sigma_0^2;y)$ is also differentiable, thus $\left.\frac{\partial l}{\partial m}\right|_{\hat{m}}=0$. Thus, when $n\to\infty$, the likelihood equation has a solution $\hat{m}$ with probability 1, and $|\hat{m}-m|<\delta$. Based on the arbitrariness of $\delta$, it follows that $\hat{m}$ is consistent with $m_0$.

Theorem 1 proves that the likelihood equation for parameter $m$ in the MVD-WLS algorithm has a solution, and the maximum likelihood estimator $\hat{m}$ is consistent. Next, we will prove that D is asymptotically normal. First, we will prove two lemmas related to regularity:

**Lemma 2:** Suppose Assumptions 2 hold, within the neighborhood of the actual parameter value $m_0$, $\frac{\partial \ln p}{\partial m},\frac{\partial^2 \ln p}{\partial m^2},\frac{\partial^3 \ln p}{\partial m^3}$ exists for all $x$, and $\left|\frac{\partial^3 \ln p}{\partial m^3}\right|\leq H(x),\ E(H(x))<\infty.$

**Proof**

Within the neighborhood of the actual parameter value $m_0$,

$$\frac{\partial \ln p}{\partial m}=-\frac{1}{2}\ln w+\frac{(y-\mathbf{x}^T\boldsymbol{\beta}_0)^2 \ln w}{2\sigma_0^2 w^m},$$

$$\frac{\partial^2 \ln p}{\partial m^2}=-\frac{(y-\mathbf{x}^T\boldsymbol{\beta}_0)^2(\ln w)^2}{2\sigma_0^2 w^m},$$

$$\frac{\partial^3 \ln p}{\partial m^3}=\frac{(y-\mathbf{x}^T\boldsymbol{\beta}_0)^2(\ln w)^3}{2\sigma_0^2 w^m},$$

the first three partial derivatives exist for all $x$ in the domain, since they are all compositions of elementary functions.

Consider the absolute value of the third-order partial derivative:

$$\left|\frac{\partial^3 \ln p}{\partial m^3}\right|=\frac{(y-\mathbf{x}^T\boldsymbol{\beta}_0)^2 |\ln w|^3}{2\sigma_0^2 w^m},$$

according to Assumptions 2, we can obtain $\frac{1}{w^m}\leq\frac{1}{c_w^m}$, and $\ln c_w\leq \ln w\leq \ln C_w$. Therefore, there exists a finite constant $L=\max\{|\ln c_w|,|\ln C_w|\}$ such that C $|\ln w|^3\leq L^3$.

According to the normality assumption of the linear regression model: $y-\mathbf{x}^T\boldsymbol{\beta}_0\sim N(0,\sigma_0^2 w^m)$,

then $E(y-\mathbf{x}^T\boldsymbol{\beta}_0)^2 = \sigma_0^2 w^m \leq \sigma_0^2 C_w^m$. Therefore, there exists a control function $H(x) = \dfrac{(y-\mathbf{x}^T\boldsymbol{\beta}_0)^2 L^3}{2\sigma_0^2 c_w^m}$,

then $\left|\dfrac{\partial^3 \ln p}{\partial m^3}\right| \leq H(x)$ and:

$$E(H(x)) = \dfrac{L^3}{2\sigma_0^2 c_w^m} E(y-\mathbf{x}^T\boldsymbol{\beta}_0)^2 \leq \dfrac{L^3 C_w^m}{2c_w^m} < \infty.$$

**Lemma 3:** At the point where the parameter takes the value $m_0$:

$$E_{m_0}\left(\dfrac{p'(y;m_0,\boldsymbol{\beta}_0,\sigma_0^2)}{p(y;m_0,\boldsymbol{\beta}_0,\sigma_0^2)}\right) = 0,\ E_{m_0}\left(\dfrac{p''(y;m_0,\boldsymbol{\beta}_0,\sigma_0^2)}{p(y;m_0,\boldsymbol{\beta}_0,\sigma_0^2)}\right) = 0,$$

$$I(m_0) = E_{m_0}\left(\dfrac{p'(y;m_0,\boldsymbol{\beta}_0,\sigma_0^2)}{p(y;m_0,\boldsymbol{\beta}_0,\sigma_0^2)}\right)^2 > 0.$$

**Proof**

Since $y-\mathbf{x}^T\boldsymbol{\beta}_0 \sim N(0,\sigma_0^2 w^m)$, then $E\left[\dfrac{(y-\mathbf{x}^T\boldsymbol{\beta}_0)^2}{\sigma_0^2 w^{m_0}}\right] = 1$, consequently

$$E_{m_0}\left(\dfrac{p'(y;m_0,\boldsymbol{\beta}_0,\sigma_0^2)}{p(y;m_0,\boldsymbol{\beta}_0,\sigma_0^2)}\right) = E_{m_0}\left(\dfrac{\partial \ln p}{\partial m}\right) = -\dfrac{1}{2}\ln w + \dfrac{\ln w}{2} E_{m_0}\left[\dfrac{(y-\mathbf{x}^T\boldsymbol{\beta}_0)^2}{\sigma_0^2 w^m}\right] = 0.$$

$$\begin{aligned}\dfrac{\partial^2 p}{\partial m^2} &= \dfrac{\partial}{\partial m}\left(\dfrac{\partial p}{\partial m}\right)\\ &= \dfrac{\partial}{\partial m}\left(p(y;m,\boldsymbol{\beta}_0,\sigma_0^2)\cdot\dfrac{\partial \ln p}{\partial m}\right)\\ &= \dfrac{\partial p}{\partial m}\cdot\dfrac{\partial \ln p}{\partial m} + p(y;m,\boldsymbol{\beta}_0,\sigma_0^2)\cdot\dfrac{\partial^2 \ln p}{\partial m^2}\\ &= p(y;m,\boldsymbol{\beta}_0,\sigma_0^2)\cdot\left(\dfrac{\partial \ln p}{\partial m}\right)^2 + p(y;m,\boldsymbol{\beta}_0,\sigma_0^2)\cdot\dfrac{\partial^2 \ln p}{\partial m^2}\\ &= p(y;m,\boldsymbol{\beta}_0,\sigma_0^2)\cdot\left[\left(-\dfrac{1}{2}\ln w + \dfrac{(y-\mathbf{x}^T\boldsymbol{\beta}_0)^2 \ln w}{2\sigma_0^2 w^m}\right)^2 - \dfrac{(y-\mathbf{x}^T\boldsymbol{\beta}_0)^2 (\ln w)^2}{2\sigma_0^2 w^m}\right].\end{aligned}$$

Let $Z = \dfrac{(y-\mathbf{x}^T\boldsymbol{\beta}_0)^2}{\sigma_0^2 w^{m_0}}$, then

$$\dfrac{p''(y;m_0,\boldsymbol{\beta}_0,\sigma_0^2)}{p(y;m_0,\boldsymbol{\beta}_0,\sigma_0^2)} = \left(-\dfrac{1}{2}\ln w + \dfrac{\ln w}{2} Z\right)^2 - \dfrac{(\ln w)^2}{2} Z = \dfrac{(\ln w)^2}{4}(Z-1)^2 - \dfrac{(\ln w)^2}{2} Z,$$

since $y-\mathbf{x}^T\boldsymbol{\beta}_0 \sim N(0,\sigma_0^2 w^m)$, then $Z \sim \chi^2(1)$, thus

$$E(Z) = 1,\ E(Z^2) = Var(Z) + [E(Z)]^2 = 3,$$

$$E(Z-1)^2 = E(Z^2) - 2E(Z) + 1 = 2.$$

then
$$E_{m_0}\left(\frac{p''(y;m_0,\boldsymbol{\beta}_0,\sigma_0^2)}{p(y;m_0,\boldsymbol{\beta}_0,\sigma_0^2)}\right) = \frac{(\ln w)^2}{4}E(Z-1)^2 - \frac{(\ln w)^2}{2}E(Z) = 0.$$

Therefore $I(m_0) = E_{m_0}\left(\frac{p'(y;m_0,\boldsymbol{\beta}_0,\sigma_0^2)}{p(y;m_0,\boldsymbol{\beta}_0,\sigma_0^2)}\right)^2 = \left(-\frac{1}{2}\ln w + \frac{\ln w}{2}Z\right)^2 = \frac{(\ln w)^2}{4}E(Z-1)^2 = \frac{(\ln w)^2}{2} > 0.$

**Theorem 2:** The optimal linear combination exponent estimator $\hat{m}_n$ for heteroscedasticity obtained by the MVD-WLS algorithm exhibits asymptotic normality, i.e.:
$$\sqrt{n}(\hat{m}_n - m_0) \xrightarrow{L} N(0, I^{-1}(m_0)),$$
where $I^{-1}(m_0)$ is the Fisher information at the actual value $m_0$ of the parameter.

**Proof**

By Theorem 1, when $n \to \infty$, the likelihood equation has a solution $\hat{m}_n$ with probability 1, and $\hat{m}_n \xrightarrow{P} m_0$. Perform a Taylor expansion of $\frac{\partial l}{\partial m}$ at $m_0$:
$$\frac{\partial l}{\partial m} = \left.\frac{\partial l}{\partial m}\right|_{m_0} + (m-m_0)\left.\frac{\partial^2 l}{\partial m^2}\right|_{m_0} + \frac{(m-m_0)^2}{2}\left.\frac{\partial^3 l}{\partial m^3}\right|_{m_1},$$

where $m_1$ is between $m$ and $m_0$, since $\left.\frac{\partial l}{\partial m}\right|_{\hat{m}_n} = 0$, let $m = \hat{m}_n$,

$$0 = \left.\frac{\partial l}{\partial m}\right|_{\hat{m}_n} = \left.\frac{\partial l}{\partial m}\right|_{m_0} + (\hat{m}_n - m_0)\left.\frac{\partial^2 l}{\partial m^2}\right|_{m_0} + \frac{(\hat{m}_n - m_0)^2}{2}\left.\frac{\partial^3 l}{\partial m^3}\right|_{m_1}, \quad (9)$$

where $m_1$ is between $m$ and $m_n$, thus $m_1 \xrightarrow{P} m_0$ ($n \to \infty$). From equation (9), we obtain:
$$\sqrt{n}(\hat{m}_n - m_0) = \frac{-\sqrt{n} \cdot \frac{1}{n} \cdot \left.\frac{\partial l}{\partial m}\right|_{m_0}}{\frac{1}{n}\left\{\left.\frac{\partial^2 l}{\partial m^2}\right|_{m_0} + \frac{\hat{m}_n - m_0}{2}\left.\frac{\partial^3 l}{\partial m^3}\right|_{m_1}\right\}}, \quad (10)$$

Since $l(m, \boldsymbol{\beta}_0, \sigma_0^2; y) = \sum_{i=1}^{n} \ln p(y_i; m, \boldsymbol{\beta}_0, \sigma_0^2)$, then
$$\left.\frac{\partial l}{\partial m}\right|_{m_0} = \sum_{i=1}^{n} \left.\frac{\partial \ln p(y_i; m, \boldsymbol{\beta}_0, \sigma_0^2)}{\partial m}\right|_{m_0}, \quad \left.\frac{\partial^2 l}{\partial m^2}\right|_{m_0} = \sum_{i=1}^{n} \left.\frac{\partial^2 \ln p(y_i; m, \boldsymbol{\beta}_0, \sigma_0^2)}{\partial m^2}\right|_{m_0},$$

they are the sum of independent and identically distributed random variables. By Lemma 3, we have:
$$E_{m_0}\left(\left.\frac{\partial \ln p}{\partial m}\right|_{m_0}\right) = E_{m_0}\left(\frac{p'(y_i; m_0, \boldsymbol{\beta}_0, \sigma_0^2)}{p(y_i; m_0, \boldsymbol{\beta}_0, \sigma_0^2)}\right) = 0,$$

$$Var_{m_0}\left(\left.\frac{\partial \ln p}{\partial m}\right|_{m_0}\right) = E_{m_0}\left(\frac{p'(y_i; m_0, \boldsymbol{\beta}_0, \sigma_0^2)}{p(y_i; m_0, \boldsymbol{\beta}_0, \sigma_0^2)}\right)^2 = I(m_0),$$

According to the Central Limit Theorem: $\sqrt{n} \cdot \frac{1}{n} \cdot \left.\frac{\partial l}{\partial m}\right|_{m_0} \xrightarrow{L} N(0, I(m_0)),$

Moreover, $E_{m_0}\left(\left.\dfrac{\partial^2 \ln p}{\partial m^2}\right|_{m_0}\right) = E_{m_0}\left[\dfrac{p''(y_i;m_0,\boldsymbol{\beta}_0,\sigma_0^2)}{p(y_i;m_0,\boldsymbol{\beta}_0,\sigma_0^2)} - \left(\dfrac{p'(y_i;m_0,\boldsymbol{\beta}_0,\sigma_0^2)}{p(y_i;m_0,\boldsymbol{\beta}_0,\sigma_0^2)}\right)^2\right] = -I(m_0),$

according to the Law of Large Numbers:

$$\dfrac{1}{n}\left.\dfrac{\partial^2 l}{\partial m^2}\right|_{m_0} \xrightarrow{a.s.} -I(m_0), \tag{12}$$

$m_1$ is within the neighborhood of $m_0$ when $n \to \infty$, by Lemma 2, there exists a control function $H(x)$ satisfying $\left|\dfrac{\partial^3 \ln p}{\partial m^3}\right| \leq H(x)$, therefore $\dfrac{1}{n}\left.\dfrac{\partial^3 l}{\partial m^3}\right|_{m_1} = O_P(1)$, then

$$\dfrac{1}{n}\dfrac{\hat{m}_n - m_0}{2}\left.\dfrac{\partial^3 l}{\partial m^3}\right|_{m_1} \xrightarrow{P} 0, \tag{13}$$

from Equations (12) and (13), we see that the denominator of Equation (10) converges in probability to $-I(m_0)$. From Equation (11), we see that the numerator of Equation (10) converges in distribution to $N(0, I(m_0))$, by Slustky's theorem, Theorem 2 is proved.

## 3.4 Evaluation Criteria

This paper compares univariate-dependent weighted least squares with multivariate-dependent weighted least squares under heteroscedasticity through extensive simulation experiments and empirical analysis. The evaluation employs the following three metrics:
(i) Absolute deviation between estimated regression coefficients and theoretical values:

$$|bias_i| = \dfrac{1}{R}\sum_{k=1}^{R}\left|\hat{\beta}_i^{(k)} - \beta_i\right|, \quad i = 0, 1, 2, \cdots, p.$$

(ii) Mean Square Error between Estimated and Theoretical Regression Coefficients:

$$MSE_i = \dfrac{1}{R}\sum_{k=1}^{R}\left(\hat{\beta}_i^{(k)} - \beta_i\right)^2, \quad i = 0, 1, 2, \cdots, p.$$

(iii) Mean Absolute Error between Regression Model Predictions and Actual Values:

$$MAE_y = \dfrac{1}{Rn}\sum_{k=1}^{R}\sum_{i=1}^{n}\left|\hat{y}_i^{(k)} - y_i^{(k)}\right|,$$

where R denotes the number of simulation iterations, and n represents the sample size. $\beta_i$ denotes the theoretical value of the regression coefficient for the i-th explanatory variable, $\hat{\beta}_i^{(k)}$ denotes the estimated value of $\beta_i$ during the kth simulation, $\hat{y}_i^{(k)}$ denotes the actual value of the response variable at the kth time step, $\hat{y}_i^{(k)}$ denotes the predicted value of the response variable by the regression model at the kth iteration.

Evaluation metrics $|bias_i|$ and $MSE_i$ measure the accuracy and stability of parameter estimates under a given estimation method. $MAE_y$ measures the closeness between the model predictions obtained from the estimation method and the true values. The closer these three metrics are to 0, the more effective the estimation method is.

## 4 Simulation

This section designed three distinct forms of heteroscedasticity variation. Using R programming to generate simulated data, we compared the univariate-dependent weighted least squares estimation (denoted as M1) with the multivariate-dependent weighted least squares estimation (denoted as M2) under different sample sizes.

Let the multivariate linear regression model be:
$$y_i = \beta_0 + \beta_1 x_{i1} + \beta_2 x_{i2} + \varepsilon_i, \quad i = 1, 2, \cdots, n,$$

where $\beta_0 = 10$, $\beta_1 = 15$, $\beta_2 = 5$, $\varepsilon_i \sim N(0, \sigma_i^2)$, $\sigma_i^2 = 0.01 f(x_{i1}, x_{i2})$, the distribution of the independent variable's values is as follows: $x_{i1} \sim U(5,15)$, $x_{i2} \sim Exp(1)$. The sample size n was set to 30, 60, and 90 respectively. The three different variations in the variance of the random error $\sigma_i^2$ are as follows:

(i) $\sigma_i^2 = 0.01(x_{i1} + 3x_{i2})^2$;  (ii) $\sigma_i^2 = 0.01 x_{i1}^2$;  (iii) $\sigma_i^2 = 0.01(x_{i1} + 3x_{i2} + x_{i1}x_{i2})^2$;

Three forms of heteroscedasticity variation, combined with different sample values, yielded a total of 9 simulation experiments. Each simulation was repeated R=100 times. For each generated simulation dataset, both M1 and M2 were used to establish multivariate linear regression models. The accuracy and stability of parameter estimation for both methods were evaluated by comparing two metrics: absolute deviation and mean squared error. Simulation results are presented in Tables 1–3. The predictive performance of both methods was evaluated by comparing the mean absolute error between fitted and actual values, as illustrated in Figure 1. The effectiveness of the proposed method under various forms of heteroscedasticity variation was assessed by comparing the combination coefficients and optimal exponents of M2 across different scenarios. The computational results are presented in Table 4.

**Table 1** Comparison of methods when the variance of the random error is $\sigma_i^2 = 0.01(x_{i1} + 3x_{i2})^2$

| Sample size | Coefficient | M1 | | M2 | |
|---|---|---|---|---|---|
| | | Absolute deviation | MSE | Absolute deviation | MSE |
| n=30 | $\beta_0$ | 0.0271 | 1.2243 | 0.0215 | 0.6836 |
| | $\beta_1$ | 0.0073 | 0.0185 | 0.0074 | 0.0104 |
| | $\beta_2$ | 0.0117 | 0.3162 | 0.0112 | 0.2083 |
| n=60 | $\beta_0$ | 0.0599 | 0.5357 | 0.0365 | 0.3000 |
| | $\beta_1$ | 0.0076 | 0.0077 | 0.0041 | 0.0047 |
| | $\beta_2$ | 0.0071 | 0.2192 | 0.0047 | 0.1204 |
| n=90 | $\beta_0$ | 0.0502 | 0.3420 | 0.0269 | 0.1303 |
| | $\beta_1$ | 0.0055 | 0.0050 | 0.0039 | 0.0024 |
| | $\beta_2$ | 0.0212 | 0.1676 | 0.0180 | 0.0537 |

**Table 2** Comparison of methods when the variance of the random error is $\sigma_i^2 = 0.01 x_{i1}^2$

| Sample size | Coefficient | M1 | | M2 | |
|---|---|---|---|---|---|
| | | Absolute deviation | MSE | Absolute deviation | MSE |
| n=30 | $\beta_0$ | 0.0112 | 0.2320 | 0.0018 | 0.2210 |
| | $\beta_1$ | 0.0040 | 0.0040 | 0.0028 | 0.0040 |
| | $\beta_2$ | 0.0035 | 0.0368 | 0.0043 | 0.0373 |
| n=60 | $\beta_0$ | 0.0061 | 0.0861 | 0.0022 | 0.0983 |
| | $\beta_1$ | 0.0031 | 0.0014 | 0.0025 | 0.0016 |
| | $\beta_2$ | 0.0166 | 0.0116 | 0.0144 | 0.0140 |
| n=90 | $\beta_0$ | 0.0136 | 0.0684 | 0.0183 | 0.0710 |
| | $\beta_1$ | 0.0002 | 0.0011 | 0.0001 | 0.0011 |
| | $\beta_2$ | 0.0042 | 0.0067 | 0.0109 | 0.0069 |

**Table 3** Comparison of methods when the variance of the random error is $\sigma_i^2 = 0.01(x_{i1} + 3x_{i2} + x_{i1}x_{i2})^2$

| Sample size | Coefficient | M1 | | M2 | |
|---|---|---|---|---|---|
| | | Absolute deviation | MSE | Absolute deviation | MSE |
| n=30 | $\beta_0$ | 0.1145 | 8.0541 | 0.1059 | 2.3581 |
| | $\beta_1$ | 0.0163 | 0.1159 | 0.0018 | 0.0357 |
| | $\beta_2$ | 0.2026 | 8.5636 | 0.0006 | 0.5065 |
| n=60 | $\beta_0$ | 0.2472 | 5.8672 | 0.0375 | 1.1707 |
| | $\beta_1$ | 0.0188 | 0.0666 | 0.0220 | 0.0365 |
| | $\beta_2$ | 0.1084 | 3.1009 | 0.1447 | 1.5534 |
| n=90 | $\beta_0$ | 0.1487 | 2.0604 | 0.0085 | 0.7099 |
| | $\beta_1$ | 0.0158 | 0.0343 | 0.0157 | 0.0127 |
| | $\beta_2$ | 0.0481 | 2.0942 | 0.0118 | 0.7124 |

**Table 4** Comparison of combination coefficients and optimal exponents results in MVD-WLS under three scenarios

| Case | $k_2/k_1$ | | | m | | |
|---|---|---|---|---|---|---|
| | n=30 | n=60 | n=90 | n=30 | n=60 | n=90 |
| (i) | 3.4174 | 2.9848 | 3.0483 | 1.9741 | 2.0121 | 1.9928 |
| (ii) | 0.2434 | 0.1553 | 0.0732 | 1.9816 | 1.9843 | 1.9951 |
| (iii) | 6.3862 | 6.1445 | 8.8771 | 2.2507 | 2.2253 | 2.1039 |

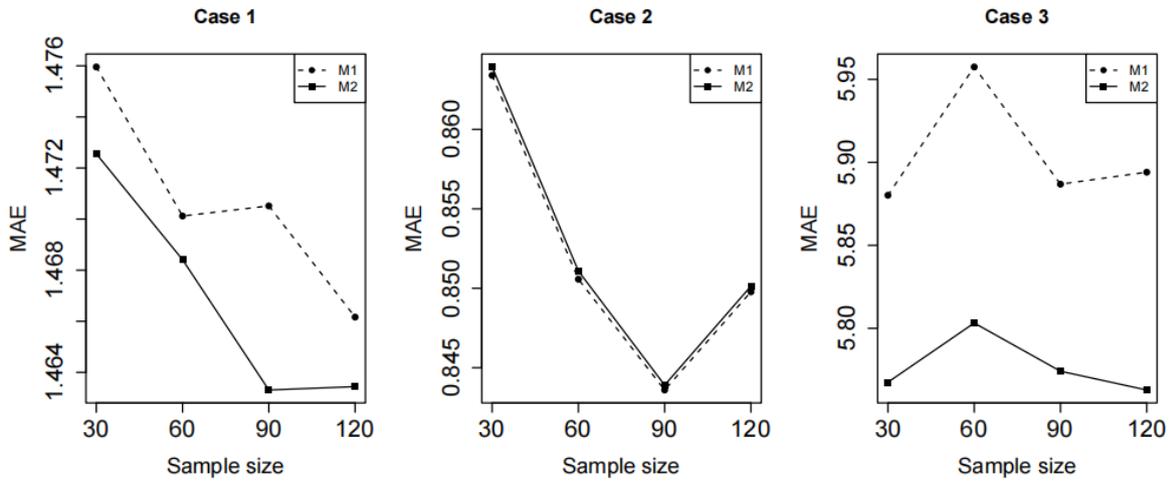

**Fig. 1** The line-plot for comparison of MAE between methods M1 and M2 under three scenarios

In Scenario 1, when the variance of random error is related to both explanatory variables, Table 1 shows that under different sample sizes, M2 outperforms M1 in both parameter estimation accuracy and stability, with the gap gradually widening as the sample size increases. The corresponding combination coefficients and optimal exponents calculations in Table 4 align closely with theoretical specifications, indicating that M2 can accurately estimate the form of heteroscedasticity variation. As shown in Figure 1, M2 also delivers superior predictive performance to M1 across various sample sizes.

In Scenario 2, when the the variance of random error is solely related to one explanatory variable, theoretically constructing the weighting function only requires selecting the explanatory variable most strongly correlated with the error variance variation. Thus, choosing M1 to estimate the heteroskedasticity model is reasonable in this scenario. As shown in Table 2, the parameter estimation accuracy of M2 differs little from M1 across different sample sizes and even outperforms M1 in some cases. The results for combination coefficients and optimal exponents in Table 4 indicate that M2 remains effective in estimating the form of heteroscedasticity variation. The MAE variation curves for both methods in Figure 1 are nearly identical. Taken together, these analyses suggest that when the variation in the error variance is indeed solely related to a single explanatory variable, M2 can naturally degenerate into M1.

In Scenario 3, when the variance structure of the random error is associated with multiple explanatory variables and exhibits complex patterns, Table 3 shows that the parameter estimates of M1 deviate significantly from theoretical expectations. In contrast, M2 maintains good accuracy and stability, and Table 4 demonstrates its adaptability under complex heteroscedasticity. As shown in

Figure 1, the predictive performance of M2 is markedly superior to that of M1. Based on the above analysis, it is evident that under complex heteroscedasticity, the estimation and predictive capabilities of M1 exhibit certain limitations. In contrast, M2 maintains relatively stable accuracy, demonstrating its adaptability to complex scenarios.

## 5 Real Data Applications

To further demonstrate the accuracy and effectiveness of the method proposed in this paper, this section employs two distinct sets of empirical data to conduct a comparative analysis from multiple perspectives. The analysis examines the performance of the univariate-dependent weighted least squares estimation (denoted as M1) versus the multivariate-dependent weighted least squares estimation (denoted as M2).

### 5.1 China Consumer Spending Influencing Factors Data

Data sourced from the National Bureau of Statistics of China website, employing an economic census methodology to compile per capita household consumption expenditure $y$ among urban residents across China's 31 regions, along with three influencing factors: income $x_1$, food expenditure $x_2$, and expenditure on entertainment and cultural services $x_3$. Establish a preliminary multivariate linear regression model: $y_i = \beta_0 + \beta_1 x_{i1} + \beta_2 x_{i2} + \beta_3 x_{i3} + \varepsilon_i$, $i = 1, \cdots, 31$. The White test revealed significant heteroscedasticity in the model, and Spearman's rank correlation analysis yielded the same conclusion. The Spearman rank correlation coefficients and corresponding p-values between each explanatory variable and the absolute residuals are shown in the table below:

Table 5  Results of Spearman's rank correlation test

| Explanatory variable | Spearman's rank correlation coefficient | p-value |
|---|---|---|
| $x_1$ | 0.4670 | 0.0081 |
| $x_2$ | 0.4585 | 0.0094 |
| $x_3$ | 0.5020 | 0.0040 |

As shown in Table 5, all explanatory variables exhibit strong and comparable correlations with the absolute residual values. Selecting only the variable with the highest correlation ($x_3$) to estimate heteroskedasticity may yield suboptimal results. We then estimated the regression model parameters using two approaches: univariate-dependent weighted least squares estimation (M1) and multivariate-dependent weighted least squares estimation (M2). The mean absolute error (MAE) and relative standard error (RSE) were calculated for the models obtained from M1 and M2. The results are presented in Table 6:

Table 6  Results of model parameter estimation and evaluation metric

| Method | $\hat{\beta} = (\hat{\beta}_0, \hat{\beta}_1, \hat{\beta}_2, \hat{\beta}_3)^T$ | MAE | RSE |
|---|---|---|---|
| M1 | $(0.2158, 0.4961, 0.3363, 1.4829)^T$ | 0.2283 | 1.0830 |
| M2 | $(0.2210, 0.5034, 0.3254, 1.4290)^T$ | 0.2157 | 0.9424 |

From the perspective of MAE, the regression model based on M2 exhibits a smaller mean absolute error and outperforms M1 in predictive performance. Zhang Xiaoqin also utilized this empirical dataset when proposing a two-stage estimation method for heteroscedastic models [29], comparing her approach with Zhang Heguang's grouped two-stage estimation [28] using MAE as the evaluation metric. This paper builds upon that work and further demonstrates that the model derived from M2 achieves the relatively smallest MAE. From the perspective of RSE, the precision of M2's sample estimates is higher than that of M1. The smaller RSE makes the model more accurate and reliable when applied to interval estimation and interval forecasting of new values.

## 5.2 Boston Housing Price Data

This dataset originates from the Boston housing price data within the R language's MASS package. It contains information on housing prices in the Boston area and serves as a classic dataset for regression analysis.

Establish a multivariate linear regression model: $y_i = \beta_0 + \beta_1 x_{i1} + \cdots + \beta_{13} x_{i13} + \varepsilon_i$, $i = 1, \cdots, 506$, no multicollinearity was detected based on the variance inflation factors of the explanatory variables. Using stepwise regression, 11 explanatory variables were selected for subsequent regression analysis. The White test indicated significant heteroscedasticity in the model. A 5-fold cross-validation approach was employed to compare the parameter estimation performance of M1 and M2, with the specific steps outlined below:

**Step 1: Dataset partitioning**

Randomly split the dataset into training and test sets in a 5:5 ratio

**Step 2: Model training and evaluation**

Two distinct multivariate linear regression models were established using the training set via M1 and M2, respectively. The performance of both models was evaluated using the test set, with the sum of squared errors (SSE) between the fitted values and actual values calculated for each model on the test set.

**Step 3: Repeated cross-validation**

Repeat the above steps 100 times, calculate and compare the mean residual sum of squares for both models.

Using R programming to implement the cross-validation process described above, the final average residual sum of squares results for the two models are as follows:

**Table 7**  Comparison of Mean SSE Across Model Cross-Validation

| Mean SSE of M1 | Mean SSE of M1 |
|---|---|
| 7005.316 | 6458.341 |

The results of cross-validation indicate that the regression model estimated using multivariate-dependent weighted least squares exhibits a smaller mean SSE and maintains superior predictive performance across different partitions of the dataset, demonstrating stronger generalization capabilities and greater stability. To visually illustrate the fitting quality of models estimated by both methods, the following plot compares actual and fitted values of the housing price data:

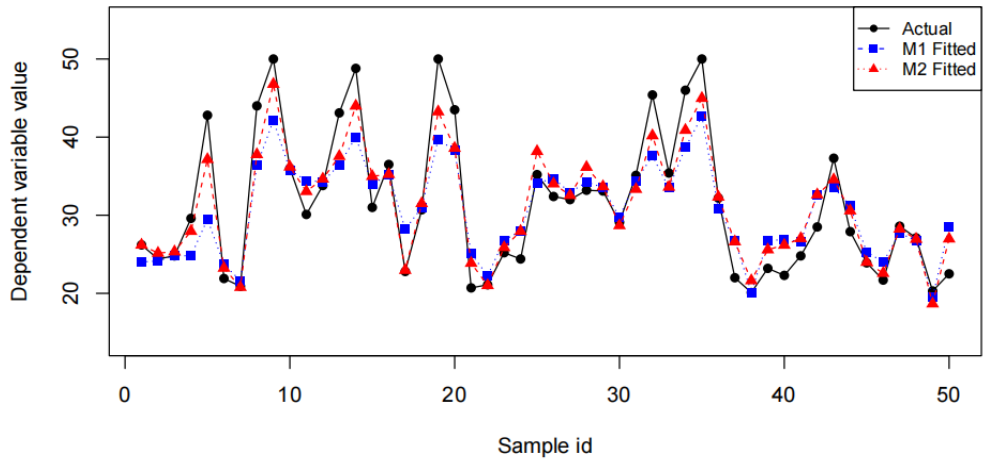

**Fig. 2** The line-plot for comparison of response variable values with fitted values from two methods

  Due to the large sample size, Figure 2 shows a segment of the overall comparison curve where the response variable exhibits significant fluctuations, illustrating a local fitting scenario. It is clearly visible that M1 demonstrates relatively poor fitting performance during periods of substantial data fluctuations, revealing certain limitations. In contrast, M2's fitted line better captures the actual data fluctuations, maintaining high accuracy even in areas with pronounced volatility.

## 6 Conclusion

  Heteroscedasticity in multivariate linear regression models is often jointly induced by multiple explanatory variables. To address the limitations in weight function construction when applying weighted least squares estimation with individual explanatory variable weights, this paper proposes a multivariate-dependent weighted least squares (MVD-WLS) estimation. This method constructs linear combinations of explanatory variables, maximizes the Spearman's rank correlation coefficient between these combinations and the absolute residuals, and employs maximum likelihood estimation for heteroskedasticity. It more comprehensively reflects the variance trends in the error term, thereby enhancing the accuracy and reliability of the regression model. This paper demonstrates that the optimal linear combination's power function estimator for heteroscedasticity fluctuations obtained by the MVD-WLS algorithm possesses consistency and asymptotic normality. In simulation experiments, this paper designed three forms of heteroscedasticity variation. Comparing the performance of univariate-dependent weighted least squares estimation and the MVD-WLS method across different sample sizes—evaluated through metrics such as estimation bias, MSE, and MAE—the results demonstrate that the MVD-WLS method achieves greater accuracy in both parameter estimation and model prediction. In real data analysis, two real-world datasets were used to compare the two methods from different perspectives. Using consumer spending as an example, the MVD-WLS method demonstrated superior precision in model forecasting and interval estimation through MAE and RSE metrics. Using Boston housing prices as another example, we examined cross-validation and model fit to demonstrate the MVD-WLS method's stable generalization capability across different data partitions, along with the relative advantage of its models in fitting data with significant volatility.

  Regarding correlation metrics, this paper employs Spearman's rank correlation coefficient to characterize the degree of association between variables and residual magnitudes. Future research may explore replacing Spearman's rank correlation coefficient with the maximum information coefficient or distance correlation coefficient to more comprehensively capture nonlinear or non-monotonic heteroscedasticity patterns, thereby enhancing the method's adaptability in complex data scenarios.

## Acknowledgement

This research is supported by the National Natural Science Foundation of China, General Program, Grant No. 12571312.